\documentclass[twocolumn,preprintnumbers,amsmath,amssymb,nofootinbib]{revtex4}

\usepackage{graphicx}
\usepackage{dcolumn}
\usepackage{bm}
\usepackage{subfigure}
\usepackage{inputenc}

\begin{document}

\title{Self Tuning Scalar Fields in Spherically Symmetric Spacetimes}

\author{${}^{a}$Stephen Appleby}
\affiliation{${}^{a}$Asia Pacific Center for Theoretical Physics, Pohang, Gyeongbuk 790-784, Korea}

\date{\today}

\begin{abstract}
We search for self tuning solutions to the Einstein-scalar field equations for the simplest class of `Fab-Four' models with constant potentials. We first review the conditions under which self tuning occurs in a cosmological spacetime, and by introducing a small modification to the original theory - introducing the second and third Galileon terms - show how one can obtain de Sitter states where the expansion rate is independent of the vacuum energy. We then consider whether the same self tuning mechanism can persist in a spherically symmetric inhomogeneous spacetime. We show that there are no asymptotically flat solutions to the field equations in which the vacuum energy is screened, other than the trivial one (Minkowski space). We then consider the possibility of constructing Schwarzschild de Sitter spacetimes for the modified Fab Four plus Galileon theory. We argue that the only model that can successfully screen the vacuum energy in both an FLRW and Schwarzschild de Sitter spacetime is one containing `John' $\sim G^{\mu}{}_{\nu} \partial_{\mu}\phi\partial^{\nu}\phi$ and a canonical kinetic term $\sim \partial_{\alpha}\phi \partial^{\alpha}\phi$. This behaviour was first observed in \cite{Babichev:2013cya}. The screening mechanism, which requires redundancy of the scalar field equation in the `vacuum', fails for the `Paul' term in an inhomogeneous spacetime. 
\end{abstract}

\maketitle % --------------------------------------------------------------------

\section{\label{sec:1}Introduction}

Regardless of whether one accepts the standard $\Lambda$CDM cosmological model or believes that the observed late time acceleration is due to some exotic and as yet unknown dynamical energy component \cite{Copeland:2006wr}, the cosmological constant problem remains an open issue. The magnitude of the vacuum energy $\rho_{\Lambda}$ is generically fixed by the UV cut-off of the underlying effective field theory in the matter sector. However the vacuum energy will gravitate, and cosmological constraints on its magnitude can be placed. The $\sim {\cal O}(10^{60})$ order of magnitude discrepancy between the value of $\rho_{\Lambda}$ allowed by cosmological observations and its particle physics expectation value can only be ameliorated by fine tuning the `bare' value of the cosmological constant appearing in the Einstein Hilbert action. However, this fine tuning is not stable and must be repeated at each order when calculating loop contributions to the vacuum. Such an instability is an indicator that the actual value of $\rho_{\Lambda}$ is sensitive to the full UV completion of the effective field theory in the matter sector (see refs.\cite{Weinberg:1988cp,Martin:2012bt} for a thorough review).

There have been a number of novel approaches to resolving this issue - see refs.\cite{Nobbenhuis:2004wn,Padilla:2015aaa,ArkaniHamed:2002fu,Kaloper:2013zca,Kaloper:2014dqa,Kaloper:2014fca, Charmousis:2011bf,Charmousis:2011ea} for a non-exhaustive list. In this work we focus on an interesting recent proposal in which an attempt is made to dynamically cancel the effect of the vacuum energy using a scalar field non-trivially coupled to the spacetime curvature \cite{Charmousis:2011bf,Charmousis:2011ea}. This effectively renders the cosmological constant problem moot, as we no longer care about the value that $\rho_{\Lambda}$ takes since it will not gravitate. The starting point is the Horndeski action - the most general covariant scalar tensor action that gives rise to second order field equations. This action is passed through a theoretical filter, in which three conditions are imposed on the underlying theory - that Minkowski space should be a solution to the field equations regardless of the value of the cosmological constant, that this solution should persist through a discontinuous change to the vacuum energy, and that the equations should admit non-trivial dynamics away from the Minkowski vacuum solution. After applying these conditions to the Horndeski Lagrangian, the authors arrived at the following action \cite{Charmousis:2011bf,Charmousis:2011ea}

\begin{widetext}
\begin{equation} S_{\rm FF} = \int \sqrt{-g} d^{4}x \left[ V_{\rm John}(\phi) G^{\mu\nu}\nabla_{\mu}\phi \nabla_{\nu}\phi + V_{\rm Paul}(\phi) P^{\mu\nu\alpha\beta} \nabla_{\mu}\phi \nabla_{\alpha}\phi \nabla_{\nu}\nabla_{\beta}\phi + V_{\rm George}(\phi) R + V_{\rm Ringo}(\phi) \hat{G} \right] + S_{\rm mat} \end{equation}
\end{widetext}

\noindent where $G^{\mu\nu}$ is the Einstein tensor, and $\hat{G}$ and $P^{\mu\nu\alpha\beta}$ are the Gauss-Bonnet scalar and double dual of the Riemann tensor respectively

\begin{eqnarray} \nonumber & & \hat{G} = R^{\mu\nu\alpha\beta}R_{\mu\nu\alpha\beta} - 4 R^{\mu\nu}R_{\mu\nu} + R^{2} \\
\nonumber & & P^{\mu\nu}{}_{\alpha\beta} = -R^{\mu\nu}{}_{\alpha\beta} + 2 R^{\mu}{}_{[\alpha} \delta^{\nu}_{\beta]} - 2 R^{\nu}{}_{[\alpha} \delta^{\mu}_{\beta]} - R \delta^{\mu}_{[\alpha} \delta^{\nu}_{\beta]} \end{eqnarray}

\noindent The four potentials $V_{\rm J,P,G,R}$ (here abbreviated) are arbitrary functions of the scalar field $\phi$. 

The theory was originally designed to screen the cosmological constant in an FLRW spacetime. An important property of any attempt to screen the vacuum energy is that the mechanism must be applicable in both a cosmological and an inhomogeneous spacetime, such as the Schwarzschild metric. The cosmological constant will gravitate not only cosmologically but also locally, introducing a modification to the Newtonian potentials leading to potentially observable effects on the motion of celestial bodies (for example, see ref.\cite{Martin:2012bt} for a nice review). The resulting upper bound on $\rho_{\Lambda}$, while considerably weaker than the cosmological one, remains extremely small relative to particle physics scales. 

The persistence of the self tuning mechanism in the `Fab Four' model in non-vacuum spacetimes is not trivial, and in this work we investigate the conditions under which the metric potentials in a spherically symmetric spacetime remain independent of the magnitude of the vacuum energy, and how we have to modify the original theory to preserve the screening property. The work will proceed as follows. In section \ref{sec:2} we write the covariant Fab Four field equations and discuss the original self tuning mechanism expounded in \cite{Charmousis:2011bf,Charmousis:2011ea}. In section \ref{sec:dS1} we generalize the model, introducing Galileon terms and show that such an introduction can give rise to screened de Sitter vacuum states as opposed to Minkowski space. In section \ref{sec:SS} we search for spherically symmetric solutions to the field equations in which the metric potentials are independent of the cosmological constant, considering two Fab Four terms individually. We summarize our results in section \ref{sec:dis}.

\section{\label{sec:2}Field equations} 

We restrict ourselves to the simplest version of the Fab Four that can give rise to self tuning solutions for an FLRW spacetime. Specifically, we take as our starting point the following action, 

\begin{widetext}
\begin{equation}\label{eq:ff1} S = \int \sqrt{-g} d^{4}x \left[ {R \over 16\pi G} + {c_{\rm J} \over M^{2}} G^{\mu\nu}\nabla_{\mu}\phi \nabla_{\nu}\phi  +  {c_{\rm P} \over M^{5}} P^{\mu\nu\alpha\beta} \nabla_{\mu}\phi \nabla_{\alpha}\phi \nabla_{\nu}\nabla_{\beta}\phi   - {\rho_{\Lambda} \over 2} \right]  + S_{\rm mat} \end{equation}
\end{widetext}

\noindent That is, we fix $V_{\rm George} = {\rm constant} = 1/16\pi G$, $V_{\rm John} = {\rm constant} = c_{\rm J}/M^{2}$ and $V_{\rm Paul} = {\rm constant} = c_{\rm P}/M^{5}$. $c_{\rm J,P}$ are dimensionless constants and $M$ is the mass that fixes the strong coupling scale of the scalar field. Since we are considering the simplest case in which the four potentials are constant, the Gauss-Bonnet contribution (Ringo) will simply reduce to a boundary term. 

The scalar field and Einstein equations read

\begin{widetext}
\begin{eqnarray} \label{eq:st1} & & {2 c_{\rm J} \over M^{2}} G^{\mu}{}_{\nu} \nabla_{\mu}\nabla^{\nu} \phi + {3 c_{\rm P} \over M^{5}} P^{\mu\nu}{}_{\alpha\beta} \nabla_{\mu}\nabla^{\alpha}\phi \nabla_{\nu}\nabla^{\beta}\phi + {3 \over 8} {c_{\rm P} \over M^{5}} \hat{G} \nabla_{\alpha}\phi \nabla^{\alpha}\phi = 0 \\
 \nonumber & & 
G^{\mu}{}_{\nu} = 8\pi G \rho_{\Lambda}g^{\mu}{}_{\nu} + {c_{\rm J} \over M^{2} M_{\rm pl}^{2}} \left[ G^{\mu}{}_{\nu} \nabla_{\alpha}\phi \nabla^{\alpha}\phi - 2 P^{\mu\alpha}{}_{\nu\beta} \nabla_{\alpha}\phi \nabla^{\beta}\phi + {1 \over 2} \delta^{\mu\alpha\beta}_{\nu\sigma\delta} \nabla^{\sigma}\nabla_{\alpha}\phi \nabla^{\delta}\nabla_{\beta}\phi \right] + \\ 
& & \hspace{6mm} + {c_{\rm P} \over M^{5} M_{\rm pl}^{2}} \left[ {3 \over 2} \nabla_{\lambda}\phi \nabla^{\lambda}\phi P^{\mu\alpha}{}_{\nu\beta}\nabla_{\alpha}\nabla^{\beta}\phi + {1 \over 2} \delta^{\mu\alpha\beta\gamma}_{\nu\delta\sigma\theta}\nabla_{\alpha}\nabla^{\delta}\phi \nabla_{\beta}\nabla^{\sigma}\phi \nabla_{\gamma}\nabla^{\theta}\phi \right]   \end{eqnarray}
\end{widetext}

\noindent where 

\begin{equation} \delta^{\mu\alpha\beta ... }_{\nu\gamma\sigma ...} = \left| \begin{array}{cccc}
\delta^{\mu}_{\nu} & \delta^{\mu}_{\gamma} & \delta^{\mu}_{\sigma} & ... \\
\delta^{\alpha}_{\nu} & \delta^{\alpha}_{\gamma} & \delta^{\alpha}_{\sigma} & ...  \\
\delta^{\beta}_{\nu} & \delta^{\beta}_{\gamma} & \delta^{\beta}_{\sigma}   & ... \\
 ... & ... & ... & ... \end{array}\right| \end{equation}

\noindent and we have introduced $M_{\rm pl}^{2} = 8\pi G$. The Fab Four scalar field equation ($\ref{eq:st1}$) has a particular property that allows the theory to screen an arbitrary cosmological constant $\rho_{\Lambda}$. To observe the mechanism, let us write the scalar field and Friedmann equations for an FLRW metric 

\begin{widetext}
\begin{eqnarray} & & ds^{2} = -dt^{2} + a^{2} \delta_{ij} dx^{i} dx^{j} \\ 
\label{eq:sf2} & &  {6c_{\rm J} H^{2} \over M^{2}} \left[\ddot{\phi} + 3 H \dot{\phi} \right] + {12 c_{\rm J} \over M^{2}}\dot{H} H \dot{\phi} - {9c_{\rm P}H^{2} \dot{\phi} \over M^{5}} \left[ 2H \ddot{\phi} + 3 H^{2} \dot{\phi}  + 3 \dot{H} \dot{\phi} \right]   = 0  \\
\label{eq:fr2} & & 3H^{2}= 8\pi G \rho_{\Lambda} + {9 c_{\rm J} \over M_{\rm pl}^{2} M^{2}} H^{2} \left(\dot{\phi}\right)^{2} - {15 c_{\rm P} \over M_{\rm pl}^{2} M^{5}} H^{3} \left(\dot{\phi}\right)^{3} \end{eqnarray}
\end{widetext}

\noindent where we assume spatial flatness throughout this work for simplicity. There are two important qualities associated with the scalar field equation ($\ref{eq:sf2}$). The first is that it trivially vanishes on approach to the Minkowski vacuum state, in which $\dot{H}, H^{2} \to 0$. As the equation is redundant at the vacuum solution, we must use the Friedmann equation to determine the dynamics of $\phi$ at this point. The second is that it contains time derivatives of $H$, allowing for a dynamical approach to the vacuum. The final condition imposed on the Horndeski action - that the `self tuning' solution persists during a piece-wise continuous change in the vacuum energy, ensures that the Minkowski vacuum solution is an attractor. 

If we define a vacuum solution as $\dot{H}=0$, $\ddot{\phi}=0$, then there are two vacuum states that solve equations ($\ref{eq:sf2},\ref{eq:fr2}$) - $\dot{\phi} = 0$, $H = H_{0}$ and $H = 0$, $\dot{\phi} = \alpha_{0}$, where $\alpha_{0},H_{0}$ are constants. The former case is the standard General Relativistic vacuum, where $H^{2} = H_{0}^{2} = 8\pi G \rho_{\Lambda}/3$ is fixed by the Friedmann equation and $\dot{\phi}=0$ solves ($\ref{eq:sf2}$). The latter solution is a vacuum state regardless of the magnitude of $\rho_{\Lambda}$, with the scalar field $\dot{\phi}$ expectation value at the vacuum determined by the Friedmann equation. We note that on approach to the Minkowski fixed point, $\dot{\phi}$ will approach a constant value and so $\phi$ remains dynamical.

\subsection{\label{sec:dS1}de Sitter Self Tuning Solutions}

The original Fab Four model was designed to possess a Minkowski vacuum state. However, one can modify the theory slightly and search for different vacua, demanding only that the expansion rate of the spacetime is independent of the cosmological constant. To this end, let us write the scalar field and Friedmann equations for a slightly modified variant of the Fab Four, namely

\begin{widetext}
\begin{equation}\label{eq:ff2} S = \int \sqrt{-g} d^{4}x \left[ {R \over 16\pi G} + {c_{\rm J} \over M^{2}} G^{\mu\nu}\nabla_{\mu}\phi \nabla_{\nu}\phi  + c_{\rm 2} \nabla_{\alpha}\phi\nabla^{\alpha}\phi  - {\rho_{\Lambda} \over 2} \right]  + S_{\rm mat} \end{equation}
\end{widetext}

\noindent For an FLRW metric, the relevant equations take the form

\begin{eqnarray}\label{eq:sf3} & &  \left({6c_{\rm J} H^{2} \over M^{2}} - 2c_{2} \right) \left[\ddot{\phi} + 3 H \dot{\phi} \right] + {12 c_{\rm J} \over M^{2}}\dot{H} H \dot{\phi}    = 0  \\
& & 3H^{2}= 8\pi G \rho_{\Lambda} + {9 c_{\rm J} \over M_{\rm pl}^{2} M^{2}} H^{2} \left(\dot{\phi}\right)^{2} + {c_{2} \over M_{\rm pl}^{2}} \left(\dot{\phi}\right)^{2} \end{eqnarray}

\noindent Once again, there are two vacuum solutions to this system of equations. Setting $\dot{H}=\ddot{\phi} =0$, the two vacua are the standard GR case in which $\dot{\phi} = 0$, $H_{0}^{2} = 8\pi G \rho_{\Lambda}/3$ and $H_{0}^{2}=c_{2} M^{2}/(3c_{\rm J})$, $\dot{\phi} = \alpha_{0}$, where $\alpha_{0}$ is an unimportant constant obtained from the Friedmann equation. Now, we have two de Sitter vacuum states - the standard General Relativistic one and a second, in which $H_{0}$ is completely independent of $\rho_{\Lambda}$. This de Sitter solution was discussed in \cite{Gubitosi:2011sg}, and see also \cite{Appleby:2011aa,Appleby:2012rx}. 

One can perform the same trick with the $c_{\rm P}$ term, if we again modify the Fab Four action slightly. Taking as our starting point the following action

\begin{widetext}
\begin{equation}\label{eq:ff3} S = \int \sqrt{-g} d^{4}x \left[ {R \over 16\pi G} +   {c_{\rm P} \over M^{5}} P^{\mu\nu\alpha\beta} \nabla_{\mu}\phi \nabla_{\alpha}\phi \nabla_{\nu}\nabla_{\beta}\phi + {c_{3} \over M^{3}} \nabla_{\alpha}\phi \nabla^{\alpha}\phi \Box \phi   - {\rho_{\Lambda} \over 2} \right]  + S_{\rm mat} \end{equation}
\end{widetext}

\noindent where $c_{3}$ is the cubic Galileon, we again derive the scalar field and Friedmann equations 

\begin{widetext}
\begin{eqnarray}\label{eq:sf4} & &  \left(6c_{3} - {9 c_{\rm P} H^{2} \over M^{2}} \right) \left( 2 \ddot{\phi} + 3H \dot{\phi} \right) {H\dot{\phi} \over M^{3}} + \left(6 c_{3} - {27 c_{\rm P}H^{2} \over M^{2}}\right){\left(\dot{\phi}\right)^{2} \over M^{3}} \dot{H}    = 0  \\
\label{eq:fr2m} & & 3H^{2}= 8\pi G \rho_{\Lambda}  - {15 c_{\rm P} \over M_{\rm pl}^{2} M^{5}} H^{3} \left(\dot{\phi}\right)^{3} + {6 c_{3} \over M_{\rm pl}^{2} M^{3}} H \left(\dot{\phi}\right)^{3}  \end{eqnarray}
\end{widetext}

\noindent This system of equations also has two de Sitter vacuum states - $H_{0}^{2} = 8\pi G \rho_{\Lambda}/3$, $\dot{\phi}=0$ and $H_{0}^{2} = 2c_{3}M^{2}/(3c_{\rm P})$, $\dot{\phi} = \alpha_{0}$. It is a curiosity that the Fab Four can yield de Sitter solutions when taken in conjunction with the $c_{2,3}$ Galileon terms. We consider the possibility of deriving similar solutions for the $c_{4,5}$ Galileon contributions in future work.

\section{\label{sec:SS}Self Tuning in Spherically Symmetric Spacetimes}

The original Fab Four model was concerned solely with self tuning within the context of cosmology. In this work we are interested in the possibility that the screening behaviour might also be present in a spherically symmetric spacetime. It is well known that the cosmological constant problem is not solely the concern of cosmologists. The vacuum energy will introduce a modification to the metric potentials on astronomical scales. Such a modification would, for example, effect observables such as the perihelion advance of Mercury. The resulting upper bound on $\rho_{\Lambda}$, while considerably weaker than the cosmological one, remains extremely low relative to typical particle physics scales \cite{Martin:2012bt}. Therefore the ability of the model to screen a cosmological constant must be applicable to spacetimes other than FLRW metrics if one wishes to avoid fine tuning problems. For existing work in this direction, we direct the reader to \cite{Babichev:2013cya,Charmousis:2014zaa,Charmousis:2015aya,Babichev:2015rva,Kaloper:2013vta}. 

In this section, we search for static, spherically symmetric solutions to the coupled scalar field and Einstein equations for the general metric 

\begin{equation}\label{eq:nlss} ds^{2} = -h(r) dt^{2} + {dr^{2} \over f(r)} + r^{2} \left( d\theta^{2} + \sin^{2}[\theta] d\phi^{2} \right) \end{equation}

\noindent We do not linearize in $h(r), f(r)$. We try to keep our discussion as general as possible, discussing the conditions under which a screened solution can exist. At a later stage, we take an ansatz for $f(r)$ and $h(r)$ that is a known solution to the Einstein equations with $\rho_{\Lambda}=0$, and attempt to recover the same solution if we switch on both a non-zero vacuum energy $\rho_{\Lambda} \neq 0$ and scalar field $\phi(r,t)$. We do not assume that the scalar field is static, as we anticipate that any screening mechanism must be dynamical in nature.

\subsection{\label{sec:Af}Asymptotically flat solutions with John?} 

For a central mass $M$ and vacuum exterior, the Schwarzschild metric provides a full non-linear and stable solution to the Einstein equations. We have 

\begin{equation} h(r) = f(r) = 1 - {\mu \over r} \end{equation} 

\noindent with constant $\mu = 2GM$. We now show that no Schwarzschild (or indeed any asymptotically flat) self tuning solution exists for the `John' fab four model. We stress that our statement only applies if we wish to eliminate the effect of the cosmological constant - if we relax this assumption then numerous solutions can be obtained \cite{Momeni:2014uwa,Afshordi:2014qaa,Korolev:2014hwa,Minamitsuji:2014hha, Volkov:2014ooa,Charmousis:2014mia,Kobayashi:2014eva,Cisterna:2014nua,Bravo-Gaete:2013dca, Minamitsuji:2013ura,Anabalon:2013oea,Minamitsuji:2015nca} (but see also \cite{Graham:2014ina}).

Our starting point is the action ($\ref{eq:ff2}$), with $c_{2}=0$ - let us write the scalar field equation for general $h(r), f(r)$ 

\begin{widetext}
\begin{equation} \label{eq:sf6} {3 c_{\rm J} \over M^{2}} \left[ \left( { f - 1 \over r^{2}} + {f' \over r} \right) \nabla_{0}\nabla^{0}\phi + \left({f-1 \over r^{2}} + {fh' \over hr} \right)\nabla_{r}\nabla^{r}\phi + \left( {f' \over 2r} + {fh' \over 2hr} + {fh'' \over 2h} + {f' h' \over 4h} - {f (h')^{2} \over 4h^2} \right) \left(\nabla_{\theta}\nabla^{\theta}\phi + \nabla_{\phi}\nabla^{\phi}\phi \right)\right] = 0 \end{equation}
\end{widetext}

\noindent We have already discussed the conditions for self tuning in a cosmological context - principle among them is that the scalar field equation should be redundant on the vacuum solution. If we insert a Schwarzschild ansatz into ($\ref{eq:sf6}$) the equation is trivially satisfied. So far so good, however this is not a sufficient condition for self tuning to occur in this spacetime. We have two additional hurdles to overcome - the first is that the momentum constraint equation is no longer trivial as it is for the FLRW metric. This will impose an extra condition on our solution, which our ansatz might not be able to satisfy. The second is that the scalar field acquires a radial profile which must solve the $(r,r)$ Einstein equation as well as the energy constraint. Simply demanding that the scalar field equation is identically satisfied is not sufficient to guarantee that a consistent solution exists of the form we are imposing. 

In this section we are searching for asymptotically flat solutions - to leading order as $r \to \infty$ we take $f(r),h(r) \to 1$ and find the Einstein equations reduce to 

\begin{eqnarray}\label{eq:j1} & & \dot{\phi}' = 0 \\
\label{eq:j2} & & {2c_{\rm J} \over M^{2} M_{\rm pl}^{2}} {\phi' \over r} \left( 2\phi'' + {\phi' \over r} \right) = 8\pi G \rho_{\Lambda} \\
\label{eq:j3} & &  {2c_{\rm J} \over M^{2} M_{\rm pl}^{2}} \left( - \phi'' \ddot{\phi} - \ddot{\phi}{\phi' \over r} + \phi'' {\phi' \over r} \right) = 8\pi G \rho_{\Lambda} \\
\label{eq:j4} & & {2c_{\rm J} \over M^{2} M_{\rm pl}^{2}} {\phi' \over r} \left( {\phi' \over r} - 2 \ddot{\phi} \right) = 8\pi G \rho_{\Lambda} \end{eqnarray}

\noindent This system of equations have unique solution 

\begin{equation} \label{eq:j5} \phi(r,t) = \beta \left( r^{2} - t^{2} \right) + \lambda_{0} + \lambda_{1}t \end{equation}

\noindent where 

\begin{equation} \label{eq:j100} \beta =  M M_{\rm pl}\sqrt{{ 8\pi G \rho_{\Lambda} \over 24 c_{\rm J}}} \end{equation} 

\noindent and $\lambda_{0,1}$ are arbitrary constants. This is the form that our solution must approach asymptotically, if we impose Minkowski boundary conditions. 

Without making any assumptions regarding $h(r), f(r)$, we can solve the $(tr)$ Einstein equation as 

\begin{equation}\label{eq:nl1} \phi = A(t) \exp\left[ {1 \over 2} \int \left({h' \over h} + {f - 1 \over r f}\right) d\bar{r} \right] + B(r)\end{equation}

\noindent for arbitrary functions $A(t)$ and $B(r)$ of $t$ and $r$ respectively. This expression was first derived in \cite{Babichev:2013cya}. For this solution to reduce to ($\ref{eq:j5}$) as $r \to \infty$, we must fix $A(t) = -\beta t^{2}$ and $\lambda_{1} =0$. In addition we fix $\lambda_{0} = 0$ without loss of generality. Using only the asymptotic flatness assumption and the momentum constraint equation, we have completely fixed the time dependence of $\phi(r,t)$. One free function remains - $B(r)$ - which must satisfy $B(r) \to \beta r^{2}$ at spatial infinity. 

If we insert ($\ref{eq:nl1}$) into the Einstein equations, what remains is a system of {\it ordinary} differential equations for $f(r), h(r), B(r)$. For our ansatz to be valid, all explicit time dependence must drop out of the $(r,r)$, $(t,t)$ and $(\theta,\theta)$ Einstein and scalar field equations. Our procedure will be to collect powers of $t$ in the Einstein equations, and by demanding the coefficients of these terms are zero we place constraints on the functions $h(r)$, $f(r)$, $B(r)$. A solution exists if we can completely eliminate the time dependence from the Einstein equations, and if the resulting ordinary differential equations admit a consistent solution for $h(r), f(r), B(r)$. 

To eliminate the highest power of $t$ from the Einstein equations ($\sim t^{4}$), one finds that $h(r)$ and $f(r)$ must be related according to

\begin{equation} h(r) = h_{0} \exp\left[ \int {(1-f) \over f} d\bar{r} \right] \end{equation}

\noindent Using this relation, we move to the next to leading order ($\sim t^{2}$) contributions to the $(t,t)$ and $(r,r)$ equations. Eliminating these terms requires us to fix $f=1$. We are always free to define our time coordinate such that $h_{0}=1$. The condition that all time dependence drops out of the equations forces us to fix $h(r)$ and $f(r)$ to their Minkowski limits over the whole domain. In this case, the Einstein equations collapse to ($\ref{eq:j1}-\ref{eq:j4}$) for all $r$ - this solution is simply the same Minkowski self tuning first constructed in \cite{Charmousis:2011bf,Charmousis:2011ea}. We conclude that no non-trivial self tuning solution exists if we demand that the spacetime is asymptotically flat. 

Let us review what has gone wrong. The condition of asymptotic flatness, together with the momentum constraint equation, completely fixes the time dependence of the solution. What remains are the scalar field and Einstein equations, which become differential equations for $h(r), f(r), B(r)$. For these functions to be independent of the time coordinate, all explicit time dependence must drop out of the equations. This can only be achieved by fixing the coefficients of the powers of $t$ in the Einstein equations to zero separately. However, doing so forces us to fix $h(r)=f(r)=1$, in which case our solution reduces to the known Minkowski screened solution.

It is clear that the self tuning conditions are more complicated for spacetimes with reduced symmetries - the momentum constraint places non-trivial conditions on the solution.

\subsection{\label{sec:JC}Schwarzschild de Sitter solutions with John and Canonical?}

Inspired by section \ref{sec:dS1}, let us now search for Schwarzschild de Sitter solutions for the modified action ($\ref{eq:ff2}$), which contains both `John' and a canonical kinetic term for the scalar field. We search for solutions of the form 

\begin{equation}\label{eq:SdS} h(r) =  f(r) = 1 - {\mu \over r} - \beta r^{2} \end{equation}

\noindent for arbitrary constants $\mu,\beta$. The metric potentials are no longer asymptotically flat, so we cannot use equations ($\ref{eq:j1}-\ref{eq:j4}$) to fix the form of $\phi(r,t)$ as in section \ref{sec:Af}.

The scalar field equation reads 

\begin{equation} \left( c_{\rm 2} - {3 c_{\rm J} \beta \over  M^{2}}\right) \Box \phi = 0 \end{equation}

\noindent The equation is identically satisfied if we fix $c_{2} = 3c_{\rm J}\beta/M^{2}$. This is the same condition as found in section \ref{sec:dS1}. The momentum constraint equation, after fixing $c_{2}$, is given by $\dot{\phi}'=0$ - hence our solution must have the form $\phi = \kappa(t)+\omega(r)$. Consistency of the gravitational equations forces us to choose $\kappa(t) = \kappa_{1}t$ for constant $\kappa_{1}$. This choice ensures that there is no explicit time dependence. The $(r,r)$ equation then reads 

\begin{widetext}
\begin{equation} 8\pi G \rho_{\Lambda} + {c_{\rm J} \over M_{\rm pl}^{2} M^{2}} \kappa_{1}^{2} {\left(\mu - 2\beta r^{3} \right) \over r^{3} h} + {c_{\rm J} \over M_{\rm pl}^{2} M^{2}} {h \left( -1 + 3\beta r^{2}\right) \over r^{2}} (\omega')^{2} = 3 \beta \end{equation} 
\end{widetext}

\noindent which has solution 

\begin{equation} \omega' =  {\kappa_{1} \over h} \sqrt{1-h} \end{equation}

\noindent if we also fix the constant $\kappa_{1}$ as 

\begin{equation} \kappa_{1}^{2} = {M^{2}M_{\rm pl}^{2}  \over 3 \beta c_{\rm J}} \left( 8\pi G \rho_{\Lambda} - 3\beta \right) \end{equation}

\noindent Somewhat surprisingly, this solution also satisfies the Hamiltonian constraint equation, and hence constitutes an exact solution to the Einstein and scalar field equations. This solution was first obtained in \cite{Babichev:2013cya}. 

Let us summarize. The `John' Fab Four term in isolation can give rise to self tuning Minkowski solutions cosmologically. If we include a canonical kinetic term, then we can also obtain de Sitter solutions starting from an FLRW metric. When we move to a spherically symmetric spacetime, the John term alone cannot give a self tuned Schwarzschild (or any asymptotically flat) solution. However, John with a non-zero $c_{2}$ can yield an exact Schwarzschild de Sitter solution, where the asymptotic de Sitter state is independent of the magnitude of $\rho_{\Lambda}$. `John' with a canonical kinetic term can screen the vacuum energy in both homogeneous and inhomogeneous spacetimes.

\subsection{Spherically Symmetric, Self Tuning Solutions with Paul?}

Let us now move onto `Paul' - fixing $c_{\rm J}=0$ and $c_{\rm P} \neq 0$. Initially we also set $c_{3} = 0$. We begin by showing, as for the `John' case, that no consistent, self tuned and asymptotically flat solution to the field equations exist (unless the space-time is exactly Minkowski). We begin by noting that for any metric in which $h(r), f(r) \to 1$ in the far field limit, the `Paul' Einstein equations are asymptotically of the following form

\begin{eqnarray} \label{eq:p1} & & 3c_{\rm P} {(\phi')^{2} \phi'' \over M^{5} M_{\rm pl}^{2} r^{2}} - 8\pi G \rho_{\Lambda} = 0 \\ 
\label{eq:p2} & & 3c_{\rm P}{(\phi')^{2} \ddot{\phi} \over M^{5}M_{\rm pl}^{2} r^{2}} + 8\pi G \rho_{\Lambda} = 0 \\ 
\label{eq:p3} & &  3c_{\rm P} {\phi' \phi'' \ddot{\phi} \over M^{5} M_{\rm pl}^{2} r} + 8\pi G \rho_{\Lambda} = 0 \\
\label{eq:p4} & & 3c_{\rm P} {(\phi')^{2}\dot{\phi}' \over M^{5}M_{\rm pl}^{2} r^{2}} = 0 \end{eqnarray}

\noindent where we have used ($\ref{eq:p4}$) to set $\dot{\phi}' = 0$ in ($\ref{eq:p3}$). Aside from the `trivial', GR case $\phi={\rm constant}$, $\rho_{\Lambda}=0$, this system of equations has solution 

\begin{equation} \label{eq:p5} \phi = \beta \left( r^{2} - t^{2} \right) \end{equation} 

\noindent where 

\begin{equation} \beta = \left({\pi G \rho_{\Lambda} M^{5} M_{\rm pl}^{2} \over 3 c_{\rm P}}\right)^{1/3} \end{equation}

\noindent This result is no surprise - for an asymptotically flat spacetime any self tuning mechanism must simply reduce to the Minkowski one in which $\phi = \phi(r^{2}-t^{2})$ as $r \to \infty$. 
  
Returning to the full non-linear equations, the solution to the $(tr)$ Einstein equation that satisfies ($\ref{eq:p5}$) at spatial infinity is given by 

\begin{eqnarray} & &  \nabla_{0} \nabla_{r}\phi = 0 \\ 
\label{eq:cross} & & \phi = A(t) \sqrt{h} + B(r) \end{eqnarray}

\noindent for arbitrary functions $A(t)$, $B(r)$. To satisfy the asymptotic condition ($\ref{eq:p5}$), we must set $A(t)=-\beta t^{2}$ (this function is independent of $r$, and so is completely fixed by the boundary condition). We have one remaining function, $B(r)$, which approaches $B(r) \to \beta r^{2}$ as $r \to \infty$. The scalar field and $(t,t)$, $(r,r)$, $(\theta,\theta)$ Einstein equations then reduce to a system of ordinary differential equations for $B(r), h(r), f(r)$. As in the case of `John', consistency of the solution requires that all time dependence must drop out of these equations - by setting successive powers of $t$ to zero we obtain a system of constraints that the ansatz must satisfy. The highest order time dependence appearing is now $\sim t^{6}$ - the requirement that the coefficients of these terms are zero forces us to fix $h'=0$. This also removes all $\sim t^{4}$ terms. To eliminate the $\sim t^{2}$ dependence, we must further fix $f=1$. We arrive at the same conclusion as for `John' - no asymptotically flat solution exists in which self tuning occurs, other than the original Minkowski case $f=h=1$.  

Let us now switch on the $c_{3}$ term and search for a solution of the form 

\begin{equation} f(r) = h(r) = 1-\beta r^{2} \end{equation} 

\noindent We already know that such a solution should exist - it is simply the same de Sitter state obtained for an FLRW spacetime in section \ref{sec:dS1} recast in static coordinates. Let us derive it. 

If we impose the following relationship between $c_{3}$ and $c_{\rm P}$

\begin{equation} c_{3} =  {3 \beta c_{\rm P} \over 2 M^{2}} \end{equation}

\noindent then the scalar field equation is identically satisfied. Furthermore, the $(t,r)$ Einstein equation reduces to $\dot{\phi}'=0$, with solution $\phi = \kappa(t) + \omega(r)$. One can ensure that all time dependence drops out of the remaining Einstein equations by setting $\kappa(t) = \kappa_{1} t$. What remains is a system of ordinary differential equations for $\omega(r)$. One can show that a consistent solution to the equations exists, where $\omega'(r)$ satisfies the cubic polynomial

\begin{equation} -6\beta {c_{\rm P} \over  M_{\rm pl}^{2} M^{5}} h^{2} {(\omega')^{3} \over r} + 6\beta {c_{\rm P} \over M_{\rm pl}^{2} M^{5}} \kappa_{1}^{2} {\omega' \over r}  = 8\pi G \rho_{\Lambda} - 3\beta \end{equation}

\noindent if we fix 

\begin{equation} \kappa_{1} = \left[ \left( 8\pi G \rho_{\Lambda} - 3 \beta \right){M^{5}M_{\rm pl}^{2} \over 6c_{\rm P} \beta^{3/2}} \right]^{1/3}  \end{equation} 

\noindent This serves as a useful check on our equations.

Let us finally search for Schwarzschild de Sitter solutions of the form ($\ref{eq:SdS}$), with $\beta$ and $\mu$ independent of $\rho_{\Lambda}$. Such a solution was found for `John', and thus far our calculations involving `Paul' have closely mimicked this case.  However one can show that for `Paul', one of the key properties of self tuning (redundancy of the scalar field equation) breaks down as soon as we take a non-vacuum metric ansatz. 

Let us write the scalar field equation for the action ($\ref{eq:ff3}$) as

\begin{widetext}
\begin{equation}\label{eq:fP} \left[ 3 {c_{\rm P} \over M^{2}} P^{\mu\alpha}{}_{\nu\beta} + 2c_{3} \left( \delta^{\mu}_{\nu} \delta^{\alpha}_{\beta}  - \delta^{\alpha}_{\nu} \delta^{\mu}_{\beta}\right) \right] \nabla_{\mu}\nabla^{\nu}\phi \nabla_{\alpha}\nabla^{\beta}\phi + \left( {3 \over 8} {c_{\rm P} \over M^{2}} \hat{G} \delta^{\alpha}_{\beta} - 2 c_{3} R^{\alpha}{}_{\beta} \right) \nabla_{\alpha}\phi \nabla^{\beta}\phi = 0 \end{equation} 
\end{widetext}

\noindent For a vacuum spacetime, $\hat{G}$ and all components of $P^{\mu\alpha}{}_{\nu\beta}$ and $R^{\alpha}{}_{\beta}$ are constant, and by virtue of their contraction with the symmetric tensor $\nabla^{\beta}\nabla_{\alpha}\phi$, the tensors $P^{\mu\alpha}{}_{\nu\beta}$ and $\left( \delta^{\mu}_{\nu} \delta^{\alpha}_{\beta}  - \delta^{\alpha}_{\nu} \delta^{\mu}_{\beta}\right)$ have the same symmetry properties in ($\ref{eq:fP}$). At a de Sitter point, we can write equation ($\ref{eq:fP}$) as 

\begin{widetext}
\begin{equation}\label{eq:fP2} \left[ -3 {c_{\rm P} H_{0}^{2} \over M^{2}} + 2c_{3}\right] \left( \delta^{\mu}_{\nu} \delta^{\alpha}_{\beta}  - \delta^{\alpha}_{\nu} \delta^{\mu}_{\beta}\right) \nabla_{\mu}\nabla^{\nu}\phi \nabla_{\alpha}\nabla^{\beta}\phi + \left( 9 {c_{\rm P} \over M^{2}} H_{0}^{4}  - 6 c_{3} H_{0}^{2} \right) \delta^{\alpha}_{\beta} \nabla_{\alpha}\phi \nabla^{\beta}\phi = 0 \end{equation} 
\end{widetext}

\noindent for constant $H_{0}$. The coefficients of both the $(\nabla \nabla \phi)^{2}$ and $(\nabla \phi)^{2}$ are exactly zero if we fix $c_{3}$ appropriately, for a de Sitter spacetime. However, for a Schwarzschild de Sitter metric, the function $\hat{G}$ and components of $P^{\mu\alpha}{}_{\nu\beta}$ contain an explicit radial dependence, which cannot be canceled by the constant $c_{3}$ term. Furthermore, the components of $P^{\mu\alpha}{}_{\nu\beta}$ are not equal, so we cannot write 

\begin{equation} P^{\mu\alpha}{}_{\nu\beta}  \sim A(r)\left( \delta^{\mu}_{\nu} \delta^{\alpha}_{\beta}  - \delta^{\alpha}_{\nu} \delta^{\mu}_{\beta}\right) \end{equation} 

\noindent in equation ($\ref{eq:fP}$). The coefficients of the scalar field kinetic terms $(\nabla \nabla \phi)^{2}$ and $(\nabla \phi)^{2}$ are therefore generically non-zero over the whole $r$-domain (in fact they are singular in the limit $r \to 0$), and there is no redundancy in the equations. This destroys the self tuning condition, which requires the scalar field equation to be trivially satisfied on the screened solution (`on-shell' in the language of \cite{Charmousis:2011bf,Charmousis:2011ea}). In fact on approach to the central singularity $r \to 0$, the scalar field must be fixed by the scalar field equation, independently of both $c_{3}$ and $\rho_{\Lambda}$, due to the potentially divergent $\hat{G}(\partial \phi)^{2}$ and $P^{\mu\alpha}{}_{\nu\beta}\nabla^{\nu}\nabla_{\mu}\phi \nabla^{\beta}\nabla_{\alpha}\phi$ terms at this point.   

For comparison, let us write the covariant scalar field equation for `John' plus canonical, discussed in section \ref{sec:JC}. In this case, screened Schwarzschild de Sitter solutions were obtained. The equation reads 

\begin{equation} 2\left( {c_{\rm J} \over M^{2}} G^{\mu}{}_{\nu} + c_{2} \delta^{\mu}_{\nu} \right) \nabla_{\mu}\nabla^{\nu}\phi = 0 \end{equation} 

\noindent Again, for a de Sitter state all components of $G^{\mu}{}_{\nu}$ are simply constant and $G^{\mu}{}_{\nu} \propto \delta^{\mu}_{\nu}$. The important difference between this case and `Paul' is that for a Schwarzschild de Sitter spacetime, the constancy of $G^{\mu}{}_{\nu}$ is preserved, and the field equation remains redundant. This allows screened solutions to be obtained.

\section{\label{sec:dis}Discussion} 

In this work we have searched for spherically symmetric solutions in the `Fab-Four' class of scalar-tensor field theories. As the spacetimes that we consider are not necessarily vacuum states, it is not clear {\it a priori} that the self-tuning mechanism used within a cosmological context in \cite{Charmousis:2011bf,Charmousis:2011ea} will persist. One must check whether solutions can be obtained in which the metric components are independent of the vacuum energy. 

We focused on the simplest class of screening solutions in which the Fab Four potentials reduce to constants. We have argued that no asymptotically flat solutions exist for either `Paul' or `John' in isolation, in which the metric potentials are independent of the vacuum energy. The requirement of screening coupled to asymptotic flatness forced the equations to collapse to the original Minkowski space solution of \cite{Charmousis:2011bf,Charmousis:2011ea}. 

When considering Schwarzschild de Sitter spacetimes, we reproduced the same solution as obtained in \cite{Babichev:2013cya} for `John' with a canonical term. We also found a new de Sitter solution, involving Paul and the third Galileon term. However we argued that this de Sitter state could not be promoted to a Schwarzschild de Sitter solution, as this spacetime destroys one of the key conditions for screening - that the scalar field equation is identically satisfied at the `vacuum'. 

It is clear that the condition that the screening mechanism must be applicable in an inhomogeneous spacetime further restricts the viable model space, tightening the noose on the ability of scalar-tensor theories to address the cosmological constant problem. However the Fab Four model survives, and it seems that `John' plus canonical has a special place in the pantheon of scalar-tensor theories, being able to screen a cosmological constant in both an FLRW and Schwarzschild de Sitter spacetime. 

A number of extensions to this work can be undertaken. To begin, it would be of considerable interest to test the stability properties of the screened Schwarzschild de Sitter solution. By introducing an explicit time dependence to the metric potentials, and evolving the combined system $\phi(r,t), f(r,t), h(r,t)$ from an initially perturbed state, one can deduce whether the solution is an attractor. In addition to the demand that perturbations do not grow, we should also check the boundedness of the Hamiltonian.

In addition, the author would like to introduce matter to the theory, and consider how one might construct a viable cosmology consistent with current observations \cite{Copeland:2012qf}. As stated in \cite{Linder:2013zoa}, the simplest Fab Four action has the undesirable property that it screens not only the vacuum energy but also dark matter and baryons. Understanding how the scalar field should couple to matter remains an open question. 

Finally, one would like to test the quantum stability of the screening mechanism. Although the existence of screened de Sitter solutions is of considerable interest, critics would argue that we have simply swapped one fine tuning for another - forced as we are to set the mass scale $M$ associated with the scalar field to $M \sim {\cal O} (H_{0})$, where $H_{0}$ is the observationally determined (small) vacuum energy. Understanding the stability of the scalar field kinetic terms within the context of radiative corrections is really the crux of the issue and will determine whether the Fab Four provides any improvement over the typical $\rho_{\Lambda}$ fine tuning. This is a direction of future study.

\acknowledgements{The author would like to thank Ed Copeland, Richard Battye and Eric Linder for helpful discussions. S.A.A wishes to acknowledge support from the Korea Ministry of Education, Science and Technology, Gyeongsangbuk-Do and Pohang City for Independent Junior Research Groups at the Asia Pacific Center for Theoretical Physics. The author would also like to acknowledge the support of the National Research Foundation of Korea (NRF-2013R1A1A2013795).}

\end{document}